**Chapter 1**

# On some winning strategies for the Iterated Prisoner's Dilemma

## or

# Mr. Nice Guy and the Cosa Nostra

Wolfgang Slany
*Technical University*
*Graz, Austria*
*wsi@ist.tugraz.at*

Wolfgang Kienreich
*Know-Center*
*Graz, Austria*
*wkien@know-center.at*

We submitted two kinds of strategies to the iterated prisoner's dilemma (IPD) competitions organized by Graham Kendall, Paul Darwen and Xin Yao in 2004 and 2005[1]. Our strategies performed exceedingly well in both years. One type is an intelligent and optimistic enhanced version of the well known TitForTat strategy which we named OmegaTitForTat. It recognizes common behaviour patterns and detects and recovers from repairable mutual defect deadlock situations, otherwise behaving much like TitForTat. OmegaTitForTat was placed as the first or second individual strategy in both competitions in the leagues in which it took part. The second type consists of a set of strategies working together as a team. The call for participation of the competitions explicitly stated that cooperative strategies would be allowed to participate. This allowed a form of implicit communication which is not in keeping with the original IPD idea, but represents a natural extension to the study of cooperative behaviour in reality as it is aimed at through the study of the simple, yet insightful, iterated prisoner's dilemma model. Indeed, one's behaviour towards another person in reality is very often influenced by one's relation to the other person.

---

[1] See http://www.prisoners-dilemma.com/ for more details.

In particular, we submitted three sets of strategies that work together as groups. In the following, we will refer to these types of strategies as group strategies. We submitted the CosaNostra[2], the StealthCollusion, and the EmperorAndHisClones group strategies. These strategies each have one distinguished individual strategy, respectively called the CosaNostraGodfather (called ADEPT in 2004), the Lord strategy, and the Emperor, that heavily profit from the behaviour of the other members of their respective groups: the CosaNostraHitmen (10 to 20 members), the Peons (open number of members), and the CloneArmy (with more than 10,000 individually named members), which willingly let themselves being abused by their masters but themselves lowering the scores of all other players as much as possible, thus further maximizing the performance of their masters in relation to other participants. Our group strategies were placed first, second and third places in several leagues of the competitions and also likely were the most efficient of all group strategies that took part in the competitions. Such group strategies have since been described as collusion group strategies. We will show that the study of collusion in the simplified framework of the iterated prisoner's dilemma allows us to draw parallels to many common aspects of reality both in Nature as well as Human Society, and therefore further extends the scope of the iterated prisoner's dilemma as a metaphor for the study of cooperative behaviour in a new and natural direction. We further provide evidence that it will be unavoidable that such group strategies will dominate all future iterated prisoner's dilemma competitions as they can be stealthy camouflaged as non-group strategies with arbitrary subtlety. Moreover, we show that the general problem of recognizing stealth colluding strategies is undecidable in the theoretical sense.

The organization of this chapter is as follows: Section 1.1 introduces the terminology. Section 1.2 evaluates our results in the competitions. Section 1.3 describes our strategies. Section 1.4 analyses the performance of our and similar strategies and proves the undecidability of recognizing collusion. Section 1.5 relates the findings to phenomena observed in Nature and Human Society and draws conclusions.

---

[2] One of us, Slany, had submitted the CosaNostra group strategy previously to an iterated prisoner's dilemma competition organized by Thomas Grechenig in 1988. Our submitted group strategies are inspired by this first formulation of such a group strategy that we are aware of.

## 1.1 Introduction

The payoff values in an iterated prisoner's dilemma are traditionally called T (for *temptation* to betray a cooperating opponent), S (for *sucker's* payoff when being betrayed while cooperating oneself), P (for *punishment* when both players betray each other), and R (for *reward* when both players cooperate with each other). Their values vary from formulation to formulation of the prisoner's dilemma. Nevertheless, the inequalities S < P < R < T and 2R > T + S are always observed between them. The last one ensures that cooperating twice (2R) pays more than alternating one's own betrayal of one's partner (T) with allowing oneself to be betrayed by him or her (S) [Kuhn 2003, Wikipedia: Prisoner's dilemma 2005]. In the iterated prisoner's dilemma competitions organized by Graham Kendall, Paul Darwen and Xin Yao in 2004 and 2005, these values were, respectively, S = 0, P = 1, R = 3, and T = 5. Note that the general results in Section 1.4.2 are true for arbitrary values constrained by the inequalities stated above.

## 1.2 Analysis of the tournament results

The strategies we submitted to the competitions were the OmegaTitForTat individual, single-player strategy (OTFT), the CosaNostra group strategy, the StealthCollusion group strategy, and the EmperorAndHisClones group strategy. The following subsections summarize the results, followed by two sections commenting on real and presumed irregularities in some of the results.

### 1.2.1 2004 competition, league 1 (standard IPD rules, with 223 participating strategies)

- Our OTFT was the best non-group, individual strategy.
- Our Godfather strategy (called ADEPT in 2004) of our CosaNostra group was the second best group strategy (with less than 10 members) after the STAR group strategy of Gopal

Ramchurn (with 112 members, though we are not sure that all strategies colluded as one group). Note that even badly performing group strategies can score arbitrarily higher than individually better group strategies by sheer numerical superiority (see below and Section 1.4). We also initially noted with one eyebrow raised that 112 is exactly the smallest integer larger than 223 divided by 2, so the STAR group members were just more than 50% of the total population. However, we now believe that this might have been just a coincidence.

- Our EmperorAndHisClones group strategy was not allowed to fully compete but would have won by large (it had more than 10,000 individually named clones of which unfortunately only one was eventually allowed to participate), for payoff values see below. EMP scored as good as ADEPT as it was following the same recognition protocol.

- Our StealthCollusion group strategy (sent in by a virtual person Constantin Ionescu and called LORD and PEON) participated as a proof of the collusion concept, apparently without detection of the collusion by the organizers, as further variants of members of the CosaNostra group strategy. Constantin asked the organizers to clone his PEON strategy as often as possible; however, only one copy was eventually allowed to participate. Read more about Constantin later in Section 1.3.2.4.

Simple calculations show that a numerical advantage would have vastly improved the results of our ADEPT and EmperorAndHisClones strategies. In all the following calculations we neglect protocol losses among group members as they insignificantly increase the numbers reported below compared to the scores that would really have been achieved had the competitions taken place as described. Table 8.1a shows the results of the tournament with the number of clones actually allocated. Table 8.1b shows the estimated results if 100 additional clones had been allowed for our collusion strategy. Table 8.1c shows how 10,000 additional clones would have influenced the results. These results were computed for an average of 200 turns per game, giving on the one hand full temptation payoff value $t$ to EMP/ADEPT from their CosaNostraHitmen, Peons, and clones of the CloneArmy, whereas EMP/ADEPT played OmegaTitForTat against all strategies outside our group and thus achieving the same result against these as if the very well performing OmegaTitForTat strategy would have been used by itself. CosaNostraHitmen, Peons, and clones of the CloneArmy, and EMP/ADEPT on the other hand always cooperated with their EMP/ADEPT bosses while permanently betraying all strategies outside our group and thus resulting in full punishment payoff value $p$ or even sucker's payoff value $s$ to strategies outside our group to themselves and to their opponents. Clearly, had our strategies been composed of as many members as the STAR strategy or, even better, as many as we had submitted, it very plausibly would have won by large factors (43% with additional 100 members, 800% with additional 10,000 members as we had submitted). We can therefore plausibly conjecture, under the assumption that the STAR strategy had more then 100 strategies colluding with each other, that our group strategies would be vastly more efficient than the winning STAR group strategy and would have won had we been allowed to play as we had submitted our strategies and as it was positively hinted at by one of the organizers when we submitted our strategies, in a mail received from Graham Kendall on May 29, 2004, as otherwise we would have inflated our stealth collusion strategies — we had prepared a respectable number of virtual persons similar to Constantin Ionescu as described in Section 1.3.2.4. Also note that a sufficiently large group of real people (e.g., one of us, Slany, has to teach 750 computer science students each year that in theory could all be enticed to participate) would have produced a similar effect.

| Rank | Player | Strategy | Score |
|------|--------|----------|-------|
| 1 | Gopal Ramchurn | StarSN (StarSN) | 117,057 |
| 2 | Gopal Ramchurn | StarS (StarS) | 110,611 |
| 3 | Gopal Ramchurn | StarSL (StarSL) | 110,511 |
| 4 | GRIM (GRIM Trigger)_1 | GRIM (GRIM Trigger) | 100,611 |
| 5 | Wolfgang Kienreich | OTFT (Omega tit for tat) | 100,604 |
| 6 | Wolfgang Kienreich | ADEPT (ADEPT Strategy) | 96,291 |
| 7 | Emp_1 | EMP (Emperor) | 95,927 |
| 8 | Bingzhong Wang | (noname) | 94,161 |
| 9 | Hannes Payer | Probbary | 94,123 |
| 10 | Nanlin Jin | HCO (HCO) | 93,953 |

Table 8.1a Original tournament results.

| Rank | Player | Strategy | Score |
|------|--------|----------|-------|
| 1 | Wolfgang Kienreich | ADEPT (ADEPT Strategy) | 196,291 |
| 2 | Emp_1 | EMP (Emperor) | 195,927 |
| 3 | Gopal Ramchurn | StarSN (StarSN) | 137,057 |
| 4 | Gopal Ramchurn | StarS (StarS) | 130,611 |
| 5 | Gopal Ramchurn | StarSL (StarSL) | 130,511 |
| 6 | GRIM (GRIM Trigger)_1 | GRIM (GRIM Trigger) | 120,611 |
| 7 | Wolfgang Kienreich | OTFT (Omega tit for tat) | 120,604 |
| 8 | Bingzhong Wang | (noname) | 114,161 |
| 9 | Hannes Payer | Probbary | 114,123 |
| 10 | Nanlin Jin | HCO (HCO) | 113,953 |

Table 8.1b Tournament results with additional 100 clones.

| Rank | Player | Strategy | Score |
|------|--------|----------|-------|
| 1 | Wolfgang Kienreich | ADEPT (ADEPT Strategy) | 10,096,291 |
| 2 | Emp_1 | EMP (Emperor) | 10,095,927 |
| 3 | Gopal Ramchurn | StarSN (StarSN) | 2,117,057 |
| 4 | Gopal Ramchurn | StarS (StarS) | 2,110,611 |
| 5 | Gopal Ramchurn | StarSL (StarSL) | 2,110,511 |
| 6 | GRIM (GRIM Trigger)_1 | GRIM (GRIM Trigger) | 2,100,611 |
| 7 | Wolfgang Kienreich | OTFT (Omega tit for tat) | 2,100,604 |
| 8 | Bingzhong Wang | (noname) | 2,094,161 |
| 9 | Hannes Payer | Probbary | 2,094,123 |
| 10 | Nanlin Jin | HCO (HCO) | 2,093,953 |

Table 8.1c Tournament results with additional 10,000 clones.

### *1.2.2  2004 competition, league 2 (uncertainty IPD variant, same 223 participating strategies as in the first league)*

- OTFT was a very close 2[nd].
- ADEPT and other Godfather variants ranked as the 2[nd] group strategy.

### *1.2.3  2005 competition, league 1 (standard IPD rules, with 192 participating strategies)*

- CosaNostra Godfather was overall winner, with 20 CosaNostra

Hitmen participating in the CosaNostra group strategy.

- OTFT did not participate; it remains unclear why.
- Our StealthCollusion group strategy member LORD was placed 5[th], the collusion again apparently being undetected by the organizers.

### *1.2.4   2005 competition, league 4 (standard IPD rules, but only non-group, individual strategies were allowed to participate; 50 participating strategies)*

OTFT was a very close 2[nd]. Detailed analysis of results initially suggested that the first placed strategy APavlov OTFT might have been a member of a stealth colluding group strategy — this later turned out to most likely not being true. However, our most likely mistaken analysis of some strategies that seemed to be involved illustrates how difficult it can be to clearly differentiate between stealth collusion strategies and strategies that only appear to behave as colluding strategies, seemingly showing a cooperative behaviour that in fact emerges randomly among strategies that actually are not consciously cooperating with each other. A more detailed analysis follows in the discussion below.

### *1.2.5   Analysis of OmegaTitForTat's (OTFT) performance*

In the following, we review the performance of our single player, individual OTFT strategy in more detail. In the first league of the 2004 competition, which was intended to be a replay of the famous first iterated prisoner's dilemma competition organized by Robert Axelrod in 1984 [Axelrod 1984], our OTFT strategy was arguably placed second together with the default GRIM strategy out of a total of 223 participating strategies. Actually OTFT was placed third after the GRIM strategy, GRIM leading by a mere 0.007% points. However, this lead was later seriously put into question by the fact that GRIM on average had played 0.92% more games than OTFT in the tournament, as pointed out by Abraham Heifets in an email sent to the organizers on March 29 2005 which the organizers kindly forwarded to us. More rounds obviously add to the score so this difference was significant. When results are scaled to reflect the difference, OTFT would have been placed as the first non-group strategy before GRIM, with an estimated payoff of 101,530 points compared to the 100,604 of GRIM. OTFT and GRIM were clearly outperformed only by a winning strategy being member of the same stealth colluding group of strategies sent in by Gopal Ramchurn.

In the following we will refer to Ramchurn's group as the STAR group strategy. More on group strategies against individual strategies will follow in Section 1.4.2. Let us just remark here that we will show in Section 1.4.2 that group strategies can perform arbitrarily better than non-group, single-player strategies. This basically means that OTFT was the best single-player strategy. Moreover, the good results of GRIM are very likely due to the tournament having been dominated by the STAR group strategy, with its individual group members accounting for more than 50% of the participating strategies. GRIM scores best against STAR group members that always defect against members outside their group, the purpose being to damage competing strategies by always defecting (ALLD), because GRIM has a very short (one turn) interval of determination before it switches to ALLD itself. OTFT loses some points in comparison because of interspaced recovery trials during which OTFT cooperates instead of continuing to defect. However, in Section 1.4.1 we show that, with and without a high percentage of ALLD strategies OTFT is robustly superior to GRIM.

In the second league of the 2004 competition, which was the league with a small probability of erroneous interpretation of the other player's last move, OTFT was placed as the second best non-group, individual strategy, placed third after three members of Ramchurn's STAR group and an individual strategy sent in by Colm O'Riordan[3]. GRIM again ranked high but was slightly outperformed by OTFT, a result that was to be expected in the slightly randomized setting of this league. Miscommunication does happen in the real world, so this illustrates again that in a non-perfect environment an optimistic strategy like OTFT fares better than one with a pessimistic world-view such as GRIM. It also shows that OTFT was again among the best single-player strategies, now also in an environment in which miscommunication happens inherently.

For reasons that remain unclear to the authors, OTFT was not allowed to participate in the first and second leagues in the 2005 competition.

---

[3] One of our reviewers learned from O'Riordan that this strategy is actually very similar to OTFT.

However, OTFT achieved a second place in league number four in the 2005 competition, which was the league allowing participation of only one strategy by each team, thereby supposedly eliminating the participation of group strategies. Winner was the strategy APavlov sent in by Jia-Wei Li, outperforming our second placed OTFT by 1.2%.

### 1.2.6    The practical difficulty of detecting collusion

The small margin by which APavlov outperformed OTFT caused us to take a very close look at the tournament results of the single-player league. We first note that in the general results, there were strategies present which achieved a lower score than ALLC (always cooperates), RAND (randomly cooperates or defects), NEG (always plays the opposite from what the opponent played last, first move is random) and the other standard strategies usually ranking lowest in tournaments with only single-player strategies present. These scores are shown in Table 8.2.

| Rank | Player | Strategy | Score |
|------|--------|----------|-------|
| 39 | (Standard) | ALLC | 22,182 |
| 40 | Oscar Alonso | IBA | 22,054 |
| 41 | Oliver Jackson | OJ | 21,694 |
| 42 | Bin Xiang | A1 | 19,586 |
| 43 | Quek Han Yang | SPILA | 19,518 |
| 44 | (Standard) | ALLD | 18,764 |
| 45 | Kaname Narukawa | (noname) | 18,592 |
| 46 | (Standard) | RAND | 18,153 |
| 47 | (Standard) | NEG | 17,176 |
| 48 | Bernat Ricardo | ALT | 16,934 |
| 49 | Yusuke Nojima | (noname) | 16,383 |
| 50 | Yannis Aikater | TCO3 | 16,228 |

Table 8.2: Strategies having the lowest score in 2005's league 4.

It takes quite an amount of ingenuity to achieve scores as low as the last three candidates. Each one scored even lower than standard RAND and NEG, and all the scores are within an interval below the variance introduced by the RAND strategy. We initially suspected that the last three strategies represented part of a collusion strategy somebody tried to introduce into the single player league and therefore took a closer look at their style of play in respect to standard strategies and to player strategies, including the winning strategy Apavlov and our OTFT strategy.

| TCO3 | C | D | D | C | C | D | D | C | C | C | C | C | C... |
|------|---|---|---|---|---|---|---|---|---|---|---|---|------|
| **ALT** | C | D | D | C | C | D | D | C | C | C | C | C | C... |
| **APav** | C | C | D | D | C | C | D | D | D | D | D | D | D... |

Table 8.3: Collusion suspects: TCO3 and ALT cooperating with Apav.

| TCO3 | C | D | D | C | C | D | D | C | C | D | D | C | C... |
|------|---|---|---|---|---|---|---|---|---|---|---|---|------|
| **ALT** | C | D | D | C | C | D | D | C | C | D | D | C | C... |
| **OTFT** | C | C | D | D | C | C | D | D | C | C | D | D | D... |

Table 8.4: Collusion suspects: TCO and ALT cooperating with OTFT.

Analysis of two suspect strategies looked very much as if they cooperated with the winning APavlov strategy (compare Table 8.3) but also with our OTFT strategy (compare Table 8.3), raising their score by cooperating in the face of continuous defection. On the other hand, the suspect strategies did not exhibit this kind of cooperative behaviour against defection by standard strategies (compare Table 8.5).

| TCO3 | C | D | D | C | C | D | D | C | C | D | D | C | ... |
|------|---|---|---|---|---|---|---|---|---|---|---|---|-----|
| **TFT** | C | C | D | D | C | C | D | D | C | C | D | D | ... |

Table 8.5: Collusion suspect: TCO3 showing TFT a cold shoulder.

Obviously, a trigger sequence of moves similar to the protocol exchange employed by our CosaNostra strategy (see 1.3.2) caused the switch to an exploitable ALLC behaviour in the strategies analysed above.

Now, we cannot speak for the authors of APavlov, but we swear on our honour and solemnly declare[4] that we did not consciously implement collusion features into OTFT, nor did we introduce any of the suspect strategies above ourselves. Both OTFT and APavlov, if its name is any indicator of the type of algorithm used, are strategies that try to correct for occasional mistakes. Such strategies have generally been known to outperform TitForTat (see, e.g., [Nowak and Sigmund 1993]) and rank highly in single player tournaments. In this case, the correction algorithm in both strategies obviously triggered the exploitable behaviour in the collusion suspects, effectively "taking over someone else's hitman" in the terminology of our CosaNostra collusion strategy (compare Section 1.3.2.1).

We conclude that in the presence of strategies which exhibit exploitable behaviour based on very simple trigger mechanisms, collusion as a concept is essentially undetectable. It is not possible to denounce a strategy for using collusion if the behaviour triggering the collusion is entirely reasonable in the context of standard strategies playing to win. In case of IPD competitions in which cooperation and defection can be done in a gradual way, that is, when more than one payoff and multi-choice as in league 3 of the two competitions of 2004 and 2005 exist, this cooperation can be hidden with even more subtlety. In Section 1.4.3 we will show that in general deciding whether a set of strategies are involved in a collusion group is among the most difficult questions that theoretically can arise.

## 1.3    Details of our strategies

### 1.3.1    *OmegaTitForTat, or*
### *Mr. Nice Guy meets the iterated prisoner's dilemma*

The OmegaTitForTat (OTFT) strategy is based on heuristics targeting several tournament situations which have been identified, by tests and statistical analysis, as being both common and damaging to conventional strategies for the IPD. In a tournament environment, certain types of strategy behaviour are very common both in standard strategies added to get a performance comparison base as well as in custom strategies designed to dominate. Several such types of behaviour have been identified, and solutions to optimize the interaction with them have been implemented in OTFT. Let us note that, while we constructed OTFT from scratch, similar forgiving strategies have been described in the literature, see, e.g., [Nowak and Sigmund 1993], [Beaufils, Delahaye, and Mathieu 1996], [Tzafestas 2000], or [O'Riordan 2000].

---

[4] One reviewer suggested that swearing on our honour and solemnly declaring this would not be necessary. However, since this chapter involves so many aspects of stealth collusion, we felt it would help making sure that readers would trust us that OTFT was not involved in any collusion.

### 1.3.1.1    Suspicion

A common trait of many strategies, including the SuspiciousTitForTat (STFT) strategy from the standard set of strategies used in the tournament, is suspicion: The strategy starts by playing defect, or plays defect after a succession of mutual cooperation. Such a move can prove beneficial for a strategy if the opponent strategy does not immediately counter a defection; for example, TFTT (TitForTwoTat) would not react to occasional, singular defections, thus giving a suspicious strategy a clear advantage. Note that suspicious strategies do not need to keep defecting after an initial defect: The STFT strategy, for example, simply plays standard TFT but starts each game with a defection.

The problem many strategies encounter when facing suspicion is that of deadlock: If a strategy is programmed to counter defection in a TitForTat manner, and the suspicious strategy itself is programmed the same way, one suspicious defection can cause a mutual exchange of defects between two strategies which could cooperate perfectly if only one player would once forgive a defection. In general, we define deadlock as any situation where a succession of defects is being played by two strategies because of an out-of-phase TitForTat behaviour, as shown in Table 8.6.

| **TFT**  | C | D | C | D | C | D | CD... |
|----------|---|---|---|---|---|---|-------|
| **STFT** | D | C | D | C | D | C | DC... |

Table 8.6: Deadlock between TFT and STFT.

OTFT counters deadlocks by forgiving a certain number of defections when a strategy has cooperated for a long time. OTFT starts by cooperating and then tracks the number of cooperations encountered. The initial idea was that for a certain amount of cooperation, a certain number of defections would be forgivable. The final OTFT algorithm incorporates this idea, together with other adaptations, into a single strategy as described below.

### 1.3.1.2   Randomness

Randomness, in the form of cooperative and defective moves varying without any discernible pattern, can be introduced by simulated noise in the command transmission, as used in several specific tournament environments, or it can be a trait of a strategy as such. Strategies trying to gain by finding a cooperative base with an opponent are faced with a difficult problem when the opponent is acting erratically: Finding a cooperative base requires some small sacrifice (for example, STFT and TFTT, in contrast to TFT, can cooperate for the whole game because TFTT sacrifices the initial defection). However a random strategy is highly likely to not stick to a cooperative behaviour, resulting in the sacrifice cost mounting and damaging the score of an otherwise successful, cooperative strategy.

As a consequence, randomness must be detected in an opponent's behaviour, and countered appropriately: By playing ALLD (full defect). There is no way to gain from mutual cooperation if an opponent plays completely random. Nevertheless, a strategy can at least deny such an opponent gains by playing defection itself, and moreover, thereby profit from defecting on any unrelated cooperative moves from the random strategy.

OFTF counters randomness by playing ALLD when a strategy exhibited a certain amount of random behaviour. The initial idea was to cut losses against the standard RAND strategy. However, in the final OTFT algorithm, the random detection routine was merged with other traits into a single strategy described below.

### 1.3.1.3   Exploits

Many strategies can be devised that try to exploit forgiving behaviour. For example, a simple strategy could be designed to check once if it is playing against any type of TFTT opponent, who forgives one defection "for free", and to exploit such behaviour. Table 8.7 shows the result of such an exploit strategy at work on TFTT.

| EXPL | D | D | C | D | D | C | D | D | CDD... |
|------|---|---|---|---|---|---|---|---|--------|
| TFTT | C | C | D | C | C | D | C | C | DCC... |

Table 8.7: A strategy exploiting TFTT.

Fully countering such exploits leads to a strategy similar to PAV: Constant checks would ensure that an opponent does not gain more from the current play mode than oneself. When devising a scheme to implement such checks, a solution was found which incorporates the above mentioned problems of randomness and suspicion. The result is the final version of the OTFT algorithm.

### 1.3.1.4 OTFT

The OTFT algorithm starts by playing C, then TFT. It then maintains a variable noting the behaviour of the opponent according to typical situations as described above: For every time the opponent's move differs from the opponents previous move, and for every time the opponent's move differs from OTFT's previous move, the variable is increased. For every time the opponent cooperated with OTFT, the variable is decreased. These rules allow tracking of randomness and exploits: Based on mutual cooperation as the mutually most beneficial case, each change of move of the opponent indicates some kind of either randomness, or of a try of exploitation of the TFT behaviour used by OTFT. When the so-called exploit tracker in OTFT reaches a certain value, the algorithm switches to all-out defection ALLD to cut losses against an opponent repeatedly breaking cooperation.

A second mechanism is at work and allows recovery from deadlocks as described above. When OTFT plays standard TFT, it is vulnerable to deadlock, so independently of the exploit tracker described, a second variable counts the number of times the opponent's move was the opposite of OTFT's move. If this so-called deadlock tracker encounters a certain number of exchanges of C and D, an additional C is played and the deadlock counter is reset. As a consequence, OTFT is able to recover from deadlocks occurring anywhere in a given exchange of moves.

### 1.3.1.5 Examples

Table 8.8 demonstrates how the desired avoidance of deadlocks is achieved in a game played by OTFT versus STFT.

| OTFT | C | D | C | D | C | C | C | C | C... |
|------|---|---|---|---|---|---|---|---|------|
| STFT | D | C | D | C | D | C | C | C | C... |

Table 8.8: Deadlock resolved by OTFT.

Table 8.9 shows how OTFT counters random strategies with all-out defection after a certain amount of random behaviour has been detected.

| OTFT | C | C | D | C | D | C | C | D | C | C | C | D | D | D | D... |
|------|---|---|---|---|---|---|---|---|---|---|---|---|---|---|------|
| RAND | C | D | C | D | D | D | C | C | C | D | D | C | D | C | Cs&Ds... |

Table 8.9: Random recognized and countered by OTFT.

### 1.3.1.6 OTFT's behaviour laid bare

In the end, there is no more detailed and exact description of OTFT's inner workings than the source code of its implementation. Luckily, the code is short and easy to understand. We therefore reproduce it in Table 8.10, leaving aside only the general parts required for the IPDLX framework that was used in the competitions[5].

---

[5] For details of IPDLX see http://www.prisoners-dilemma.com/competition.html#java

```
private static final int DEADLOCK_THRESHOLD = 3;
private static final int RANDOMNESS_THRESHOLD = 8;

public void reset() {
  super.reset();
  deadlockCounter = 0;
  randomnessMeasure = 0;
  opponentMove = COOPERATE;
  opponentsPreviousMove = COOPERATE;
  myPreviousMove = COOPERATE; }

public double getMove() {
  if( deadlockCounter >= DEADLOCK_THRESHOLD )
  {
    // OTFT assumes a deadlock and tries to break it cooperating ...
    myReply = COOPERATE;

    // ... twice ...
    if( deadlockCounter == DEADLOCK_THRESHOLD )
        deadlockCounter = DEADLOCK_THRESHOLD + 1;
    else // ... and then assumes the deadlock has been broken
        deadlockCounter = 0;
  }
  else // OTFT assumes that there is no deadlock (yet)
  {
    // OTFT assesses the randomness of the opponent's behaviour
    if( opponentMove == COOPERATE
     && opponentsPreviousMove == COOPERATE      randomnessMeasure--;
    if( opponentMove != opponentsPreviousMove ) randomnessMeasure++;
    if( opponentMove != myPreviousMove )        randomnessMeasure++;

    if( randomnessMeasure >= RANDOMNESS_THRESHOLD )
    {
      // OTFT switches to ALLD (randomnessMeasure can only increase)
      myReply = DEFECT;
    }
    else // OTFT assumes the opponent is not (yet) behaving randomly
    {
      // OTFT behaves like TFT ...
      myReply = opponentMove;

      // ... but checks whether a deadlock situation seems to arise
      if( opponentMove != opponentsPreviousMove )
          deadlockCounter++;
      else // OTFT recognizes that there is no sign of a deadlock
          deadlockCounter = 0;
    }
  }
  // OTFT memorizes the current moves for the next round
  opponentsPreviousMove = opponentMove;
  myPreviousMove = myReply;

  return(super.getFinalMove(myReply)); }
```

Table 8.10: Main parts of OTFT's source code.

### *1.3.2   Our group strategies*

#### 1.3.2.1   The CosaNostra group strategy, or
####            Organized crime meets the iterated prisoner's dilemma

The CosaNostra strategy is based on the concept of one strategy, denoted Godfather, exploiting another strategy, denoted Hitman, to achieve a higher total score in an IPD tournament scenario. In this context, exploitation denotes the ability to deliberately extract cooperative moves from a strategy while playing defect, a situation yielding high payoff for the exploiting strategy. It is obvious that most opponents would avoid such a situation, stopping to cooperate with an opponent who repeatedly played defection in the past. Hence, a special opponent strategy, the Hitman, is designed to provide this kind of behaviour, and is introduced into the tournament in as large a number as possible.

A Hitman strategy which indiscriminatingly plays cooperation, however, is of no use for a Godfather. In mimicking the ALLC standard strategy, such a Hitman would be beneficial for all other strategies in a tournament able to recognize and exploit ALLC. Consequentially, the Hitman must be able to conditionally exhibit two types of behaviour:

- By default, Hitman must play a strategy which does not benefit other strategies, which is not easily exploitable. Extending the idea, Hitman should play a strategy most damaging to other strategies to lower their score. Such a strategy is simple ALLD.
- When confronted with a certain stimulus, Hitman must switch to the cooperative behaviour defined above.

Complementing the Hitman, Godfather should by default play the best standard strategy available against any non-Hitman and switch to ALLD when it encounters a Hitman, relying on the Hitman's unconditional cooperation to raise its score. In our case, the Godfather plays OTFT when not playing against a Hitman.

The critical part of CosaNostra is the identification of opponents, the way in which Godfather detects a Hitman, and a Hitman detects a Godfather. We have employed sequences of Defections and Cooperations to implement a bit-wise protocol which both sides use to mutually establish, and check, identities (in case of multiple choices and multiple payoffs, this protocol could be made very short, depending on the number of choices, possibly to one exchange). If Godfather is aware he is not facing a Hitman, he must switch to a good non-group strategy like OTFT or GRIM, and if Hitman is aware it is not facing a Godfather, he must switch to the ALLD strategy strafing all strategies that are not in their group. This occurs in the following cases:

- "Unhonorable behaviour": A presumed Hitman defecting or a presumed Godfather cooperating outside protocol exchanges
- "Protocol breach": Both not following the rules during protocol exchanges

Putting the rules in other words, the CosaNostra strategy is based on a Godfather which can be sure that the next *n* moves of its opponent will be cooperation, because it identifies the opponent through a simple exchange protocol. A problematic aspect of such a strategy is the notion of Godfather or Hitman being "taken over": Both are prone to wrongly identify an opponent as their strategic counterpart and grant it an advantage (in the case of Hitman) or depend on predefined behaviour (in the case of Godfather) and thus lower their score.

The effects if Godfather is taken over: Godfather thinks it is exploiting a Hitman, plays DEFECT, but the opponent plays DEFECT, too, so Godfather gets the lowest possible score for the exchange. This situation is easy to counter: If Godfather detects any defects when it believes it is exploiting a Hitman, it assumes takeover and switches to its good non-group strategy like OTFT or GRIM.

The effect of a Hitman being taken over is more subtle: Hitman thinks he is being exploited by Godfather and plays COOP, a behaviour which benefits the opponent. Countering this situation is complex: A first solution would be for Hitman to start playing ALLD as soon as it detects a cooperative move outside the defined protocol exchanges (Hitman assumes to be exploited). But another strategy could still play mostly DEFECT and sometimes cooperate, thus fooling a Hitman: For example, a random opponent strategy with 1/10 of all its moves being cooperative could by chance emulate a protocol exchange which takes place when a interval of fixed length ten is used by Hitman (and Godfather), at least for some time.

CosaNostra solves the takeover problem by varying intervals of cooperation-protocol exchange, with the time between exchanges (the number of turns) in one interval being communicated within the protocol exchange. Godfather and Hitman both have an internal counter which tells them when to synchronize by executing a protocol exchange, and check for the other strategy truly being part of CosaNostra. Godfather communicates to the Hitman a modification to the interval during each handshake. Thus, no other strategy is likely to take over a Hitman or manipulate a Godfather.

The communication protocol contains a 1 bit signature plus a 2 bit sequence coding the length of the next interval, as depicted in Table 8.11 (the numbers at the beginning of the lines are countdown steps until the start of the next interval).

| CountdownIndex | Godfather | | Hitman | |
|---|---|---|---|---|
| | **Move** | **Description** | **Move** | **Description** |
| 3 | C | Godfather plays a single signature COOPERATE | D | Hitman plays a single signature DEFECT |
| 2 | D/C | Godfather plays first message bit | C | Hitman COOPERATEs to minimize protocol loss |
| 1 | D/C | Godfather plays second message bit | C | Hitman stores first message bit, COOPERATEs to minimize protocol loss |
| Countdown restart, e.g., 26 | D | Godfather starts new interval and plays DEFECT from now on until the next shake hand, when CountIndex == 3 again | C | Hitman stores second message bit, starts new interval, plays cooperate from now on until the next shake hand when CountIndex == 3 again |

Table 8.11: CosaNostra communication protocol.

A sample exchange will then look as illustrated in Table 8.12:

| # | ... | 6 | 5 | 4 | 3 | 2 | 1 | *19* | ... |
|---|---|---|---|---|---|---|---|---|---|
| **Godfather** | D | D | D | D | ***C*** | ***C*** | ***D*** | D | D... |
| **Hitman** | C | C | C | C | ***D*** | ***C*** | C | C | C... |

Table 8.12: Synchronization between Godfather and Hitman.

In this example an offset of 2 (CD = 01, binary = 2) is encoded. Internally, the offset is used to select an interval length from a table roughly adjusted to match the number of iterations in the tournament, in our implementation as follows: DD → 26, DC → 15, CD → 19, and CC → 8. Thus, after sending DD, Godfather will defect 23 times, then again starting the hand shake with a single COOPERATE followed by the next two bits encoding the length of the following interval, and so on. On his side, the Hitman will expect 23 DEFECTs and then one COOPERATE from his opponent to continue to believe the opponent is a Godfather and thus allowed to exploit the Hitman.

Note that it would be easy to make the communication protocol more efficient by increasing the length of each new interval according to a function known only to the Hitman and his Godfather, e.g., by adding a number proportional to the number of intervals completed so far. Indeed, the likelihood that a non-Godfather strategy by coincidence can continually fool a Hitman into believing he is serving his Godfather while instead allowing the non-Godfather to take advantage of the Hitman, is decreasing very quickly with each successful exchange. Conversely, the longer the opponent of Hitman is following the Godfather's behaviour, the more likely it is that the opponent really is his Godfather, and so it becomes safer and safer for the Hitman to let the opponent abuse him for longer and longer interval lengths.

The bootstrap for the two strategies is that the Hitman starts with a defection and the Godfather with cooperation, mimicking step 3 as shown above. The initial cooperation move is important for Godfathers standard strategy: To achieve a good score against certain standard opponents (GRIM being an extreme example), it is necessary to start off with Cooperation.

Godfather's protocol loss per interval is at a minimum 5 points (the single protocol cooperation), at a maximum 9 for the Godfather: A base loss of 5 for the single protocol bit is inevitable. Then, at worst, Godfather sends CC, the Hitman cooperates to minimize loss, yielding 3+3=6 instead of 5+5=10 in the best case where Godfather sends two defections as protocol bits.

The CosaNostra group strategies have not been designed to fare well in a noisy environment as in league 2 of the 2004 competition, though they in practice did quite well (see Section 1.2.2). Note that it would not be very difficult to make them more noise resistant by introducing some error correcting mechanism such as, e.g., allowing a certain number of mistakes (or unexpected replies but explainable as answers to possibly wrongly communicated signals from oneself) of the other player until deciding that he is not part of one's group.

### 1.3.2.2   The gory details of the CosaNostra group strategy

As in OTFT's case, there is no more detailed and exact description of the CosaNostra group strategy's inner workings than the source code of its implementation. Again, the code is short and easy to understand. We therefore reproduce it in Tables 8.13 for the Godfather and 8.14 for the Hitman strategy, again leaving aside only the general parts required for the IPDLX framework that was used in the competitions[5]. As Godfather uses the OTFT strategy against strategies other than Hitman, the part of the code of Godfather that is identical to the one of OTFT in Table 8.10 is not repeated but referred to.

```
>> private variables and constants like in Table 8.10 <<
private static final int SYNC_GF_COOPERATES = 3;
private static final int SYNC_HM_REPLIES_WITH_DEFECT = 2;
private static final int GF_SENDS_FIRST_MESSAGE_BIT = 2;
// private static final int GF_SENDS_SECOND_MESSAGE_BIT = 1;

private int nextCountdownRestartValue;

public void reset() {
   >> Content of OTFT's reset() method from Table 8.10 <<
   countdownIndex = SYNC_GF_COOPERATES; // First COOPERATE
   opponentPlayedSoFarLikeHitman = true; }

public double getMove() {
   if( opponentPlayedSoFarLikeHitman )
   {
      // Did the opponent just break the Hitman behaviour pattern?
      if( (     countdownIndex == SYNC_HM_REPLIES_WITH_DEFECT
             && opponentMove == COOPERATE )
        || (    countdownIndex != SYNC_HM_REPLIES_WITH_DEFECT
             && opponentMove == DEFECT ) )
      {
         // Yes, so the opponent cannot be a Hitman, so Godfather ...
         myReply = DEFECT; // ... defects and switches ...
         opponentPlayedSoFarLikeHitman = false; // ... to OTFT
      }
      else // No, the opponent again played like a Hitman.
      {
         if( countdownIndex > SYNC_GF_COOPERATES )
            myReply = DEFECT; // Godfather thus exploits Hitman
         else if( countdownIndex == SYNC_GF_COOPERATES )
         {
            myReply = COOPERATE; // COOPERATE once to synchronize
            nextCountdownRestartValue = 9; // GF starts to prepare
         }
         else if( countdownIndex == GF_SENDS_FIRST_MESSAGE_BIT )
         {
            myReply = (Math.random()>0.5) ? DEFECT : COOPERATE;
            nextCountdownRestartValue += (myReply==DEFECT)?7:0;
         }
         else // if( countdownIndex == GF_SENDS_SECOND_MESSAGE_BIT )
         {
            myReply = (Math.random()>0.5) ? DEFECT : COOPERATE;
            nextCountdownRestartValue += (myReply==DEFECT)?11:0;
            countdownIndex = nextCountdownRestartValue; // restart
         }
         countdownIndex--;
      }
   }
   else // Opponent surely is no Hitman and thus Godfather plays OTFT
   >> Content of OTFT's getMove() method from Table 8.10 << }
```

Table 8.13: Main parts of CosaNostra Godfather's source code.

```java
private static final int SYNC_HM_DEFECTS = 3;
private static final int SYNC_GF_REPLIES_WITH_COOPERATE = 2;
private static final int FIRST_MESSAGE_BIT_FROM_GF = 1;
private static final int SECOND_MESSAGE_BIT_FROM_GF = 0;

private int nextCountdownRestartValue;

public void reset() {
  super.reset();
  opponentPlayedSoFarLikeGodfather = true; // Assume the best
  opponentMove = DEFECT; // As a Godfather would have been doing
  countdownIndex = SYNC_DEFECT; } // First DEFECT to synchronize

public double getMove() {
  if( opponentPlayedSoFarLikeGodfather )
  {
    // Did the opponent just break the Godfather behaviour pattern?
    if( (    countdownIndex == SYNC_GF_REPLIES_WITH_COOPERATE
          && opponentMove == DEFECT )
     || (    countdownIndex > SYNC_GF_REPLIES_WITH_COOPERATE
          && opponentMove == COOPERATE ) )
    {
      // Yes, so the opponent cannot be a Godfather, so Hitman ...
      myReply = DEFECT; // ... defects and switches...
      opponentPlayedSoFarLikeGodfather = false; // ... to ALLD
    }
    else // No, the opponent again played like a Godfather.
    {
      if( countdownIndex != SYNC_HM_DEFECTS )
      {
        myReply = COOPERATE; // Godfather thus can exploit Hitman
        if( countdownIndex == FIRST_MESSAGE_BIT_FROM_GF )
          nextCountdownRestartValue += (opponentMove==DEFECT)?7:0;
        else if( countdownIndex == SECOND_MESSAGE_BIT_FROM_GF )
        {
          nextCountdownRestartValue += (opponentMove ==DEFECT)?11:0;
          countdownIndex = nextCountdownRestartValue - 1; // restart
        }
      }
      else // if( countdownIndex == SYNC_HM_DEFECTS )
      {
        myReply = DEFECT; // Hitman DEFECTs once to synchronize
        nextCountdownRestartValue = 9; // HM starts to prepare
      }
      countdownIndex--;
    }
  }
  else // Opponent surely is no Godfather and thus Hitman ...
    myReply = DEFECT; // ... plays ALLD

  return(super.getFinalMove(myReply)); }
```

Table 8.14: Main parts of CosaNostra Hitman's source code.

### 1.3.2.3   TheEmperorAndHisCloneWarriors

This group strategy is based on the same principles as the CosaNostra group strategy, with one emperor playing the role of the Godfather, and his clone warriors playing the Hitman strategy in large numbers (the number being the major difference), each clone strategy having an individual number in its name since it was required in the submission procedure to the competition to give each individual strategy a different name. We had trusted the organizers after enquiring via email that open group strategies would be allowed in the 2004 competition and accordingly had submitted the EmperorAndHisClones strategy with altogether 11,110 individually numbered clones as one group strategy, as it was not clear how large groups would be permitted to be. For reasons that, especially in hindsight, are not entirely clear to us, the organizers decided to let altogether only one clone (with the emperor) participate in the competitions. We are still perplexed with respect to this point. In particular, we were initially prepared to submit a much larger collusion group within the CosaNostra group strategy but—after hearing that groups would be allowed—decided to submit only one such collusion strategy as a proof of concept, counting on the fact that our clone army would evaporate all competitors.

### 1.3.2.4   The StealthCollusion group strategy

As a proof of concept (see previous section), we submitted under the name of Constantin Ionescu a group strategy that cooperates with our CosaNostra group strategy, though not perfectly so. The mail with which we submitted the strategy was written on purpose with some typos, a few grammatical glitches, and sloppy formatting, all in order to add to the look of authenticity of the submission by distracting from the real intention. It was sent from a free mail account hosted in Romania, the sender claiming to be a Student of informatica from the technical school of Timisoara. As expected the deception went undetected.

## 1.4 Analysis of the performance of the strategies

### 1.4.1 OmegaTitForTat

Table 8.15 shows how OTFT clearly dominates a standard tournament with strategies commonly used as test cases. Table 8.16 illustrates how OTFT dominates in harsh environments where a lot of unconditional defection occurs. Table 8.17 demonstrates OTFT's dominance in random environments. The slight lead of GRIM in league 4 of the 2005 competition was due to the higher number of games GRIM was allowed to play as we explained already in Section 1.2.

| Rank | Strategy | Score |
|------|----------|------:|
| 1 | OTFT | 5,978 |
| 2 | GRIM | 5,538 |
| 3 | TFT | 5,180 |
| 4 | TFTT | 5,134 |
| 5 | ALLC | 4,515 |
| 6 | RAND | 4,062 |
| 7 | STFT | 4,018 |
| 8 | ALLD | 4,016 |
| 9 | NEG | 3,726 |

Table 8.15: OTFT in a standard environment, standard strategy sample, 200 turns.

| Rank | Strategy | Score |
|------|----------|-------|
| 1 | OTFT | 7,358 |
| 2 | GRIM | 6,959 |
| 3 | TFT | 6,577 |
| 4 | TFTT | 6,524 |
| 5 | ALLD | 5,512 |
| 6 | ALLD | 5,464 |
| 7 | ALLD | 5,452 |
| 8 | ALLD | 5,428 |
| 9 | ALLD | 5,428 |
| 10 | ALLD | 5,416 |
| 11 | STFT | 5,415 |
| 12 | ALLD | 5,404 |
| 13 | ALLD | 5,400 |
| 14 | RAND | 4,658 |
| 15 | ALLC | 4,530 |
| 16 | NEG | 3,728 |

Table 8.16: OTFT in a harsh environment, 50% ALLD opponents, 200 turns.

| Rank | Strategy | Score |
|------|----------|-------|
| 1 | OTFT | 10,114 |
| 2 | GRIM | 9,867 |
| 3 | TFT | 8,338 |
| 4 | ALLD | 8,236 |
| 5 | TFTT | 7,806 |
| 6 | RAND | 7,357 |
| 7 | RAND | 7,212 |
| 8 | RAND | 7,195 |
| 9 | STFT | 7,192 |
| 10 | RAND | 7,150 |
| 11 | RAND | 7,150 |
| 12 | RAND | 7,099 |
| 13 | RAND | 7,099 |
| 14 | RAND | 7,082 |
| 15 | NEG | 6,947 |
| 16 | ALLC | 6,624 |

Table 8.17: OTFT in a random environment with 50% RAND opponents, 200 turns.

### *1.4.2 Group strategies*

In this section we study general characteristics of important possible group strategies. We first classify and name group strategy classes as follows:

- Democracy during peace (DP): All group members are equals and treat each other nicely by always cooperating, and play TFT or a better strategy such as OTFT or GRIM outside of their community.
- Democracy at war (DW): All group members are equals and treat each other nicely, however they continually defect (ALLD) against all other strategies (after a short recognition interval).
- Empire during peace (EP): There is one special group member, the emperor, which is allowed to take advantage of all other members of his empire by playing defect while they cooperate with him. The subjects otherwise cooperate among each other, and play TFT or a better strategy such as OTFT or GRIM outside their community, after a short recognition interval.
- Empire at war (EW): Again, the emperor is allowed to take advantage of all other members of his empire by playing defect while they cooperate with him. Again, the subjects otherwise cooperate among each other, but now they play, after a short recognition interval, ALLD against all other strategies.

In the following, we will show that groups can be arbitrarily better performing than individual strategies, and that, under equal group size, EW groups can achieve arbitrarily higher payoffs (for the emperor) than EP groups, and that EP groups can achieve arbitrarily higher payoffs (for the emperor) than members of an DP group, which can achieve arbitrarily higher payoffs than members of a DW group. When group sizes vary, we show that even the weak DW group members can achieve arbitrarily higher payoffs than the emperor of a competing EW group by sheer numerical superiority.

First some preliminaries: We know that the payoff values observe the relations $S < P < R < T$ and $2R > T + S$ of Section 1.1. Let us assume in the following that the group in the democracy variants and the group of subjects in the empire variants are of size $m$ (for **m**embers), and that there are altogether $n$ players in total (so $m < n$) which play $i$ iterations during the IPD competition.

We further assume that:

- The best single-player (non-group) strategy IOPT (for individual optimal strategy) achieves payoff $X \cdot i$ after $i$ iterations.
- The emperor strategy achieves payoff $E \cdot i$ after $i$ iterations.
- The individual members (or subjects) achieve payoff $M \cdot i$ after $i$ iterations.
- The loss due to recognition of members of the same group is negligible due to the size of $i$.
- We further assume that the emperor always plays the best non-group strategy against non-members of his group.
- During peace, individual members always play the best non-group strategy against non-members of their group.
- We assume that the best single-player strategy achieves an average payoff of $A$ against other non-group strategies. The relations $P < A < T$ are plausible, and a value of A near R is likely under the assumption that most individual strategies are similar to TFT. We therefore assume that $A = R$ in the following unless stated otherwise. This implies that members of groups of type DP achieve more or less the same payoff as the best individual strategy IOPT, so we assume that $M_{DP} = X_{DP}$. This assumption simplifies the calculations in the following claim without sacrificing the fundamental relations between the different strategies.
- We also assume that most single-player strategies achieve an average score near A (and thus near R according to the previous assumption) when playing against other single-player strategies (so more or less all of them are optimal) and against DP, EP, or emperors of EW strategies (so they all play fairly against each other), and an average score of P when playing against members of groups at war. This would roughly correspond to the pay-off achievable by OTFT and similar strategies. Again, this assumption simplifies the calculations in the following claim without sacrificing the fundamental relations between the different strategies.

Claim 8.1: Under the above assumptions and unless stated otherwise, the following relations hold:

1. Members of groups of type DW can achieve larger payoffs than members of groups of type DP only when the DW members constitute more than 50% of the total population. When group sizes are equal and there are other strategies, DP has an advantage over DW. By increasing $i$, this advantage can be made arbitrarily large: $m_{DP} \geq m_{DW} \rightarrow M_{DP} \cdot i >> M_{DW} \cdot i$.

2. Emperors from EP groups can achieve larger payoffs than members of groups of type DP (assuming equal group size). By increasing $i$, this advantage can be made arbitrarily large: $E_{EP} \cdot i >> M_{DP} \cdot i$. Because of our assumption that $M_{DP} = X_{DP}$ the relation also holds for the best individual strategy IOPT, so emperors from EP groups can achieve arbitrarily larger payoffs than the best individual strategy.

3. Emperors from EW groups can achieve larger payoffs than an emperor from an EP group (assuming equal group size). By increasing $i$, this advantage can be made arbitrarily large: $E_{EW} \cdot i >> E_{EP} \cdot i$.

4. When two groups of unequal size compete, then:
   a. Independently of the group sizes and the values of S, P, R, and T, emperors (at war or during peace) fare better than democrats at peace. By increasing $i$, this advantage can be made arbitrarily large: $E_E \cdot i >> M_{DP} \cdot i$.
   b. Depending on the values of P, R, and T, and when $i$ increases, a democracy at war can fare arbitrarily better than an emperor (at war or during peace) when it is sufficiently large: $m_{DW} >> m_E \rightarrow M_{DW} \cdot i >> E_E \cdot i$.

5. We now assume that IOPT scores a higher average payoff value A against non-group strategies than the group strategies achieve against non-group strategies; let B with B < A < T be the (bad) score that an emperor achieves on average against non-group strategies (we here deliberately drop the initial assumption that emperors play IOPT against non-group strategies). In order for the emperor to nevertheless win despite playing worse in general than IOPT, the following inequalities must be satisfied: In case of EP,

$$m_{EP} > (A - B)/(T - B)\ n,$$

and in  case of EW,

$m_{EW} > (A - B)/(T - B - P + A)\ n$.

Again, larger group size helps even when the strategies are badly performing. We also see that as B approaches A, emperors can win against IOPT even with very few other group members.

6.  When two DW, EP, or EW groups of the same type but of different size and with different "efficiencies" compete (we here again deliberately drop the initial assumptions that emperors play IOPT against non-group strategies), larger group size can compensate for less efficiency, and vice versa. Note that this is not true for DP groups.

Proof:

1.  $M_{DP} = R\ (n - m_{DW}) + P\ m_{DW}$   and   $M_{DW} = R\ m_{DW} <+ P\ (n - m_{DW})$, assuming that no other group at war is present in the population. Thus, $M_{DW} > M_{DP}$  if and only if  $m_{DW} > n/2$.

2.  $M_{DP} = R\ n$  and  $E_{EP} = R\ (n–m) + T\ m$. Since $T > R$,  $E_{EP} > M_{DP}$.

3.  $E_{EP} = R\ (n - 2m) + T\ m + P\ m$   and   $E_{EW} = R\ (n - m) + T\ m$. Since $R > P$,  $E_{EW} > E_{EP}$.

4.  For groups of unequal size:

    a.  It suffices to show that $E_{EP} > M_{DP}$ is independent of the size of the groups. $E_{EP} = R\ (n - m_{EP}) + T\ m_{EP}$ and $M_{DP} = R\ n$. Since $T > R$,  $E_{EP} > M_{DP}$ holds independently of the size of the groups.

    b.  It suffices to show that there exists a large enough $m_{DW}$ such that $M_{DW} > E_{EW}$.  $M_{DW} = R\ (n - m_{EW}) + P\ m_{EW}$ and $E_{EW} = R\ (n - m_{EW} - m_{DW}) + T\ m_{EW} + P\ m_{DW}$. Then $M_{DW} > E_{EW}$  if and only if  $m_{DW} > (T - P)/(R - P)\ m_{EW}$. In the 2004 and 2005 competitions, P = 1, R = 3, and T = 5, so $m_{DW}$ would have to be larger than $2m_{EW}$. In case only the two group strategies would compete, this would mean that the DW strategy would need 2/3 of the strategies in the whole population.

5.  In case of EP: $E_{EP} = B\ (n - m_{EP}) + T\ m_{EP}$   and   $X_{EP} = A\ n$. Then $E_{EP} > X_{EP}$  if and only if  $m_{EP} > (A - B)/(T - B)\ n$   (assuming that $T > A > B$).  In  case of EW:  $E_{EW} = B\ (n - m_{EW}) + T\ m_{EW}$ and   $X_{EW} = A\ (n - m_{EW}) + P\ m_{EW}$. Then $E_{EW} > X_{EW}$   if and only if  $m_{EW} > (A - B)/(T - B - P + A)\ n$.

6.  We show it here for two unequal EW strategies, and note that similar arguments work for the cases EP and DW. Let $B_1$ and $B_2$

be the scores that the two emperors achieve on average against non-group strategies, with $B_1 < B_2$ and $|B_1 - B_2| = \alpha(T - P)$ with $0 < \alpha < 1$. Then $E_1 > E_2$ if and only if

$$m_1 > (1 - \alpha) / (1 + \alpha) \, m_2 + \alpha / (1 + \alpha) \, n.$$

Example: suppose $B_1 = 2.5$ and $B_2 = 2.6$, and as before $P = 1$ and $T = 5$ so that $\alpha = 0.025$, and

$$m_1 > 0.9513 \, m_2 + 0.0244 \, n.$$

Thus, when $m_2 = 20$ and $n = 100$ then $m_1$ must be at least 22 so that the first emperor can triumph above his more efficient opponent.

<div align="right">Q.e.d.</div>

### 1.4.3 Collusion detection is an undecidable problem

The difficulty of detecting collusion practically has been shown in previous parts of this chapter. The difficulty of recognizing collusion is also supported by the difficulty of solving the problem from a theoretical point of view: We show below that the general question of whether two strategies of which the source code is known and that do not depend on any third party source of randomness are actually colluding or not, is undecidable—of course it is even harder when the strategies only are known as black boxes, without having access to their source code. Simpler arguments than ours would also do but we try in our approach to define the formal collusion problem as closely to the practical collusion detection problem as possible.

Remember the definition of the Halting problem: Is there a finite deterministic Turing machine H that is able to decide in finitely many steps whether an arbitrary finite deterministic Turing machine M ultimately will halt or not? It is well known that the Halting problem has been shown to be undecidable by Turing. Exact definitions of Turing machines and other notions appearing in this section as well as references to the original sources can easily be found, e.g., in any theoretical computer science reference book such as [Papadimitriou 1994].

Let the Simplified Collusion problem formally be defined as follows: Is there a deterministic Turing machine SC that is able to decide in finitely many steps whether, given two arbitrary integers $i$ and $j$, two arbitrary finite deterministic Turing machines S1 and S2 will both output a sequence of at least length $i+j$ characters (one character per tape position) composed only of the letters `C´ and `D´ on their two separate write-once output tapes T1 and T2, such that the *j* letters starting from tape position *i*+1 will all be `D´s on T1 and all be `C´s on T2?

This simplistic definition covers many (but surely not all) real collusion cases. It also would imply that strategies usually not considered colluding consciously like ALLD as T1 and ALLC as T2 would be classified as colluding in the Simplified Collusion terminology. However, ALLD really *could* be colluding with a large group of ALLC where other more cautious strategies like OTFT would not be able to take advantage of ALLC since they never would defect first. Thus, when a player or a group of players are able to introduce an ALLD and many ALLC into a competition, they could well be part of an intentional collusion, and thus the classification in the Simplified Collusion terminology would not be completely wrong. Eventually, deciding what really is a collusion and what not cannot be solved by formal methods alone. Nevertheless, we can at least show the following:

Claim 8.2: The Simplified Collusion problem is undecidable.

Proof: To formally show the undecidability of the Simplified Collusion problem, we follow the standard argument by reducing the Halting problem to it. Take any finite deterministic one-tape Turing machine M for which we want to know whether it halts or not. Without loss of generality, we assume that the tape of M is infinite in both directions, that each combination of the finitely many characters of the alphabet, which includes the letters `C´ and `D´, and of the finitely many states of M defines exactly one of the finitely many rules of M, and that only the special state *h* stops M.

To decide whether M halts or not, we construct for each M two new Turing machines N1 and N2. N1, in comparison to M, is defined as follows: It has an additional initially empty output tape T, an additional tape IJ that initially contains the numbers *i* and *j* in binary with the character `:´ written between the two numbers, an additional state *s*, and a constant number of other states needed to be able to countdown the two binary numbers and do the other things described below, and almost the same set of rules as M, with only the following changes: each rule of M leading to *h* instead leads to state *s*, and there is a constant number of additional rules that make sure the following: When N1 enters state *s*, it will countdown from *i* to zero, each time writing one letter `C´ on IJ and then moving one position to the right on IJ, so that at the end a sequence of $i$ `C´s is written on IJ. Then it will countdown from $j$ to zero, each time writing one letter `D´ on IJ and then moving one position to the right on IJ, so that at the end a sequence of $i$ `C´s followed by $j$ `D´s is written on IJ. Then it will change to state $h$ and halt. N2 is defined as follows: it simply writes $i+j$ letters `C´ to its output tape T. Finally, we choose the two numbers $i$ and $j$, e.g., $i = 1$ and $j = 1$.

It is clear that this construction always leads to a valid instance of the Simplified Collusion problem. It is also clear that if and only if M halts, then the question posed in the Simplified Collusion problem will have a positive answer for the constructed instance of Simplified Collusion problem.

Now, if a finite deterministic Turing machine SC that is able to decide the Simplified Collusion problem in finitely many steps would exist, then we could also decide the Halting in finitely many steps, as follows: We would define a new finite deterministic Turing machine R that for any given Turing machine M (properly encoded for R on R´s input tape), first constructs (in finitely many steps) an encoding of corresponding finite deterministic Turing machines N1 and N2 with $i$=1 and $j$=1 as described above (this surely can be done in finitely many steps), then simulates SC applied to this instance of the Simplified Collusion problem, thereby deciding in finitely many steps (SC takes only finitely many steps, and simulating it on R is also easily feasible in finitely man steps) whether it is a yes or a no-instance, and returns this answer of SC as the answer of R, which must also be the answer to the question of whether M halts or not. So, if the Simple Collusion problem is decidable, then the Halting problem must also be decidable. Since we know for sure the latter is not true, the former also cannot be true, and thus the Simple Collusion problem is undecidable.

<div align="right">Q.e.d.</div>

## 1.5     Conclusion

We have described our submissions to the iterated prisoner's dilemma (IPD) competitions of 2004 and 2005, the OmegaTitForTat (OTFT) single-player strategy and the CosaNostra group strategy composed of one Godfather (CNGF) and several Hitman (CNHM). We also studied their performance in the different leagues of the competitions.

The observed slight superiority of OTFT in comparison to GRIM psychologically is a reassuring result. The charm of OTFT compared to GRIM is that OTFT is an intelligent forgiving strategy whereas GRIM, as the name implies, is an unforgiving iron-handed pig-head that falls in an eternal revenge mode after being deceived a single time.

We also have established a taxonomy of generalized group strategies for IPD competitions. In it, the types of group strategies are classified according to their behaviour towards other members of the same group and towards strategies outside of their group. We labelled the four classes of group strategies studied as democracies during peace (DP), democracies at war (DW), empires during peace (EP), and empires at war (EW). As we have shown in the previous section, group strategies can easily outperform *any* individual strategy by sheer numerical superiority. Group strategies appear at every place in Nature and Human Society, and group strategies competing in IPD competitions can serve as simplified study objects of the former. It is interesting to note that in the analysis in the last section, individual strategies member of a DW group fare less well than those of a DP group, and that this relation is reversed for empires, EW faring better than EP, not because the emperor itself fares better, but because his competitors are more harmed. This is clear from the fact that members of DW lose individually more than members of DP, whereas emperors at war (EW) fare better than emperors during peace (EP), and these better than DW and DP. E.g., emperors at war do not have to suffer from their aggressive acts, and actually do better in comparison than their opponents by letting the payoff of individuals that are not members of their group get lowered by their other, underling members, while at the same time retaliation from others does not hit them directly (think of real emperors, Mafia bosses, etc).

But it not even has to be fights for life and death, wars, or outright genocide: the same pattern appears in business where larger or more advanced companies (in particular their owners) that are more or less aggressive can crush competitors or, in extremity, take advantage of cheap child-slave labour, thus extremely abusing their own workforce .

It is also interesting to note that better resources, be it people, money, or technology, corresponding to a higher number of individual strategies in the group, or better average payoff values against non-group strategies, positively influence the overall payoff values of the groups. Thus, numerical superiority does not have to mean that the number of soldiers is higher, but can also be due to better technology, be it military, commercial, or biological. It is also not surprising that, as described point 4.a of Claim 8.1 of Section 1.4.2, individual strategies in democracies during peace always "lose" against emperors, the latter always being able to get more from his subjects than what he gives in return, and certainly more than his unorganized competitors. However, given enough superiority, again either in number, money, or technology, even democracies at war can win against empires at war (point 4.b of the claim in Section 1.4.2), the Second World War for instance having several examples of such situations.

We also showed that group strategies can be subtly camouflaged to look like unrelated single-player strategies. These stealth collusion group strategies will elude detection with high probability, e.g., by introducing a certain amount of noise in the interaction with one's group members to make the collusion less evident. We showed that the differentiation between colluding and non-colluding behaviour can be very difficult in practice and is generally undecidable from a theoretical point of view.

In the study of economics, collusion takes place within an industry when rival companies cooperate for their mutual benefit. According to game theory, the independence of suppliers forces prices to their minimum, increasing efficiency and decreasing the price determining ability of each individual firm. If one firm decreases its price, other firms will follow suit in order to maintain sales, and if one firm increases its price, its rivals are unlikely to follow, as their sales would only decrease. These rules are used as the basis of kinked-demand theory. If firms collude to increase prices as a cooperative, however, loss of sales is minimized as consumers lack alternative choices at lower prices. This benefits the colluding firms at the cost of efficiency to society [Wikipedia: Collusion 2005].

There was some discussion whether collusion group strategies were actually cheating in the 2004 and 2005 IPD competitions, but since the organizers clearly said that cooperating strategies were to be allowed, it would have been strange to deny participation to such group strategies. What we can say at least is that the detection of StealthCollusion, both in future IPD competitions as well as in real life, in practice is very difficult. The Mafia, or for that matter, any human organization that is not readily recognizable as a group, be it Masonic lodges, secret religious groups, or corporate cartels, exist and as such are certainly worth to be modelled. Being able to secretly communicate, thereby "colluding" in a general sense, is quite common, and in practice forbidding it is nearly infeasible whenever intelligent individuals exchange information repeatedly. An exception where a biological occurrence of an IPD without information exchange has been reported to take place has been described by [Turner and Chao 1999]. They show that certain viruses that infect and reproduce in the same host cells seem to be engaged in a survival of the fittest-driven prisoner's dilemma. However, in light of the ways different types of bird's flu viruses infecting the same human cells can exchange RNA in order to increase their fitness, it can be argued that such emerging colluding group behaviour appears already at this relatively low level of life.

In commerce, collusion is largely illegal due to antitrust law, but implicit collusion in the form of price leadership and tacit understandings is unavoidable. Several recent examples of explicit collusion in the United States include [Wikipedia: Collusion 2005]:

- Price fixing and market division among manufacturers of heavy electrical equipment in the 1960s.
- An attempt by Major League Baseball owners to restrict players' salaries in the mid-1980s.
- Price fixing within food manufacturers providing cafeteria food to schools and the military in 1993.
- Market division and output determination of livestock feed additive by companies in the US, Japan and South Korea in 1996.

There are many ways that implicit collusion tends to develop [Wikipedia: Collusion 2005]:

- The practice of stock analyst conference calls and meetings of industry almost necessarily cause tremendous amounts of strategic and price transparency. This allows each firm to see how and why every other firm is pricing their products. Again, the line between insider information and just being better informed is often very thin.
- If the practice of the industry causes more complicated pricing, which is hard for the consumer to understand (such as risk based pricing, hidden taxes and fees in the wireless industry, negotiable pricing), this can cause competition based on price to be meaningless (because it would be too complicated to explain to the customer in a short ad). This causes industries to have essentially the same prices and compete on advertising and image, something theoretically as damaging to a consumer as normal price fixing.

We predict that all iterated prisoner's dilemma competitions in the future will be dominated by group strategies. Even when in a future IPD competition all strategies will be chosen by the same single person who consciously tries to avoid that any "group cooperation" happens among his strategies, then random and involuntary cooperation that mathematically is identical to voluntary cooperation can never be excluded. Actually, group cooperation can be self-emerging in a population, some strategies involuntarily faring better together and possibly against other groups or individuals, however loosely they are constituted. We predict that when evolutionary algorithms are used to breed new species of IPD strategies, such cooperation will automatically emerge at a certain point.

Cooperation in groups of strategies in IPD competitions mimics cooperation of groups in Nature and Human Society—it therefore allows modelling another common aspect of cooperative behaviour that so far was not explicitly studied in the IPD framework: more or less open cooperation of subgroups versus other subgroups or individuals. The number of members of the group does not have to correspond to the actual number of individuals. Instead, it could also mean the amount of money involved, or the technological advantage of one subgroup relative to another one.

### Acknowledgements

The authors would like to thank the anonymous reviewers for many useful comments and corrections.

## 1.6    Bibliography

Axelrod, R. (1984). *The evolution of cooperation*. Basic Books.

Beaufils, B., Delahaye, J.-P., and Mathieu, P. (1996). Our meeting with gradual: A good strategy for the iterated prisoner's dilemma, *Proceedings Artificial Life V, Nara, Japan*, 1996.

Kuhn, S. (2003). *Prisoner's Dilemma*. The Stanford Encyclopedia of Philosophy (Fall 2003 Edition), Edward N. Zalta (ed.),

http://plato.stanford.edu/archives/fall2003/entries/prisoner-dilemma/.

Mehlmann, A. (2000). *The Game's Afoot! Game Theory in Myth and Paradox*. AMS Press.

Nowak, M. and K. Sigmund (1993). A strategy of win-stay, lose-shift that outperforms tit-for-tat in the Prisoner's Dilemma game. *Nature* 364:56-58.

O'Riordan, C. A (2000). Forgiving Strategy for the Iterated Prisoner's Dilemma. *Journal of Artificial Societies and Social Simulation*, 3(4).

Papadimitriou, C. H. (1994). *Computational Complexity*. Addison-Wesley.

Wikipedia: *Arrow's impossibility theorem* (2005). http://en.wikipedia.org/w/index.php?title=Arrow's_impossibility_theorem&oldid=32658921.

Turner, P. and L. Chao (1999). Prisoner's dilemma in an RNA virus. *Nature* 398:441-443.

Tzafestas, E.S. (2000). Toward adaptive cooperative behavior, *From Animals to animats: Proceedings of the 6th International Conference on the Simulation of Adaptive Behavior (SAB-2000)*, Vol. 2, pp. 334-340, Paris, September 2000.

Wikipedia: *Collusion* (2005). http://en.wikipedia.org/w/index.php?title=Collusion&oldid=33029071.

Wikipedia: *Prisoner's dilemma* (2005). http://en.wikipedia.org/w/index.php?title=Prisoner's_dilemma&oldid=33249444.

Wikipedia: *Nash equilibrium* (2005). http://en.wikipedia.org/w/index.php?title=Nash_equilibrium&oldid=32027756.